%% file: Ply-Based Appearance Modeling for Knitted Fabrics/main.tex
\title{A Practical Ply-Based Appearance Modeling for Knitted Fabrics}
\author[D. Fellner \& S. Behnke]
{\parbox{\textwidth}{\centering Zahra Montazeri$^{1,2}$, Søren Gemmelmark$^{2}$, Henrik W. Jensen$^{2}$, Shuang Zhao$^{1}$
        }\\
{\parbox{\textwidth}{\centering $^1$University of California, Irvine,
         $^2$Luxion Inc
      }
}
}

\begin{document}
	\teaser{
		\includegraphics[width=\linewidth]{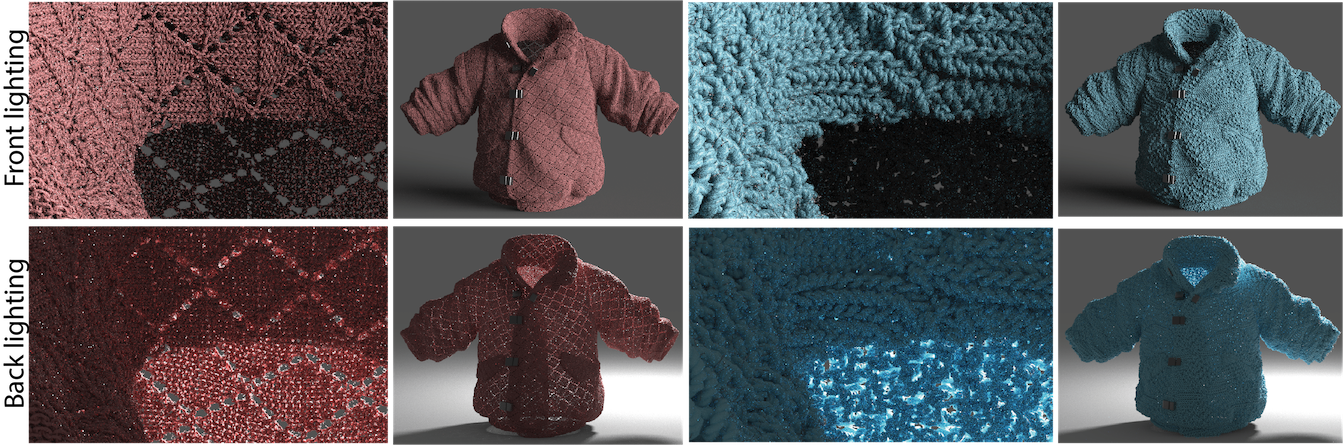}
		\centering
		\caption{
		    Our practical ply-based model for knitted fabrics can simulate both the reflection and  the transmission of light (as shown in the top and the bottom rows, respectively).
		    Additionally, we proposed an efficient technique to model the geometry of a knitted fabric with user-specified patterns. 
		}
		\label{fig:teaser}
	}
	\maketitle
	\input{abstract.tex}
	\input{intro.tex}
	\input{related.tex}

	\input{methodology}

	\input{implementation.tex}
	\input{results.tex}
	\input{conclusion}
    \printbibliography
\end{document}

%% file: abstract.tex
\begin{abstract}
    Modeling the geometry and the appearance of knitted fabrics has been challenging due to their complex geometries and interactions with light.
    Previous surface-based models have difficulties capturing fine-grained knit geometries; Micro-appearance models, on the other hands, typically store individual cloth fibers explicitly and are expensive to be generated and rendered.
    Further, neither of the models have been matched the photographs to capture both the reflection and the transmission of light simultaneously.
    
    In this paper, we introduce an efficient technique to generate knit models with user-specified knitting patterns.
    Our model stores individual knit plies with fiber-level detailed depicted using normal and tangent mapping.
    We evaluate our generated models using a wide array of knitting patterns. Further, we compare qualitatively renderings to our models to photos of real samples.
\end{abstract}

%% file: intro.tex
\section{Introduction}
\label{sec:intro}
Fabrics are essential in our daily lives. Therefore, modeling and reproducing their appearances virtually has been an active research topic in computer graphics.

Traditionally, fabrics have been modeled using surface-based models that treat a fabric an infinitely thin 2D sheet and express fabric appearance using (potentially spatially varying) reflectance models.
The surface-based models are light-weight, easy to edit, and have produced plausible results for a wide range of fabrics.
On the other hand, these models are known to have difficulties capturing a fabric's fine-grained micro-geometry that is crucial for producing realistic close-up renderings.

Alternatively, micro-appearance models depict a fabric's micron-diameter fibers individually (by utilizing high-resolution 3D volumes or fiber meshes) and offer the highest level of details to date.
Unfortunately, since these models are normally data-intensive, they are difficult to edit and expensive to render.

Recently, Montazeri~et~al.~\cite{Montazeri2020} introduced a ply-based model that bridges the gap between the surface-based and the micro-appearance models.
This model describe the geometry of a fabric as a collection of plies (where a ply is a substrand of a yarn).
The fiber details are then added to the ply surfaces using normal and tangent mapping.
Lastly, the ply-level geometry is coupled with a specialized BSDF capable of controlling the reflectance and transmittance of light simultaneously.

Despite its practicality and flexibility, Montazeri~et~al.'s ply-level model~\cite{Montazeri2020} has been designed for woven fabrics only.
Knitted fabrics---which are arguably more common than the woven ones since knitting has been the most popular method for manufacturing cloth---have been largely neglected by this model and many previous ones.
Compared with woven fabrics, knitted ones typically exhibit more three-dimensional patterns that lead to more complex scatterings of light.
Efficiently describing large pieces of knitted fabrics with fiber-level details has largely remained a challenge.

In this paper, we extends Montazeri~et~al.'s model~\cite{Montazeri2020} to handle knitted fabrics.
Concretely, our contributions include:
\begin{itemize}
    \item 
    A new pipeline that efficiently models the geometry of a knitted fabric at the ply level (Sections~\ref{subsec:flat} and \ref{ssec:transform}).
    \item An improved appearance model build upon ply-based model \cite{Montazeri2020} that better match the reference photos (Section~\ref{ssec:appearance}).
\end{itemize}
We qualitatively evaluate the accuracy of our models by comparing their renderings to photographs of physical samples with identical knitting patterns (Figure~\ref{fig:appearance}).

%% file: related.tex
\section{Related Works}
\label{sec:related}
In Computer Graphics, there are large body of works in cloth rendering and simulation. In this section, we summarize previous works related to modeling the geometry as well as the appearance of the knit fabrics.
\par\textbf{Cloth Modeling}
Most of the work involving cloth and textile has been aimed towards woven fabrics due to their simpler structure. While woven fabrics consist of two perpendicular yarns called warps and wefts, knits have usually made by one long yarn for the entire piece of the fabric. The modeling approaches for woven cloth are mostly sheet-based representation \cite{Nobuyuki2011, pENG2006}, however, recently 3D modeling approaches have also been used in cloth prototyping. These model represent the fabric down to the fiber level detail and introduced a new level of details ideal for close-up views. They are modeled either using volumetric representation with high density which is essentially for dense fabrications \cite{Zhao2011, Zhao2012}, or fiber mesh \cite{Khungurn2016} utilizing volumetric \cite{Zhao2011} or fiber-based \cite{Schroder2015}.
\par\textbf{Cloth Rendering}
In addition to the geometry modeling of the fabrics, their appearance has been another challenge and an active research. A number of surface-based  models have been developed and offer visually plausible appearance specially for far fields \cite{Irawan2012, Sadeghi2013, Adabala2003}. While these models render the results efficiently, due to their light weight base geometry, they lack the fine details specially when they are rendered from close-up views.  \\
Data-driven approaches such as Bidirectional Texture Functions (BTF) \cite{Dana1999} and specialized Bidirectional Reflectance Distribution Function (BRDF) combined with texture mapping are more commonly used to produce realistic results for fabrics. \\
Recently, a wide range of micro-appearance cloth models were introduced to the field \cite{Khungurn2016, Zhao2011, loubet2018new, Montazeri2019}. These models describe the model down to the fiber level and hence they offer a more accurate appearance thanks to their underlying geometry. However, given their complexity of their models, efficiently rendering them has been challenging hence there are several precomputation-based methods to address this problem such as Zhao et al. \cite{Zhao2013}. Khungurn et al. \cite{Khungurn2017} extended precomputed radiance transfer first proposed by Sloan et al. \cite{Sloan:2002:PRT} to caputre light transports for fabrics. These models still lack performance for an interactive setup. Also, Wu and Yuksel \cite{Wu2017} proposed a GPU-based method by utilizing precomputed `core fibers' to further optimize the performance. However, this model lacks the physical accuracy for editing and predictive rendering. Montazeri et al. \cite{Montazeri2020} proposed an efficient appearance modeling for ply-based cloth representation that accurately captures reflection as well as the transmission of light. We slightly extended this model by improving the fitting step to match the reference photo better and used for the rendered results in this paper.

%% file: methodology.tex
\section{Our Method}
\label{sec:method}
We now describe our main pipeline for generating and rendering knitted fabrics in ply-level.
Our ply-level representation is based on the work by Montazeri et al \cite{Montazeri2020} that offers a good balance between efficiency and fidelity.
Also, our approach simulates only the macro-scale base mesh then map the ply-level representation afterward.
Our method for generating the knit geometry consists of two main stages: first generating the knit curves in a flat pieces in 3D texture space, then warp them to the 3D object by extruding the yarns out of the base geometry.

\begin{figure*}[t]
	\centering
	\includegraphics[width=0.99\textwidth]{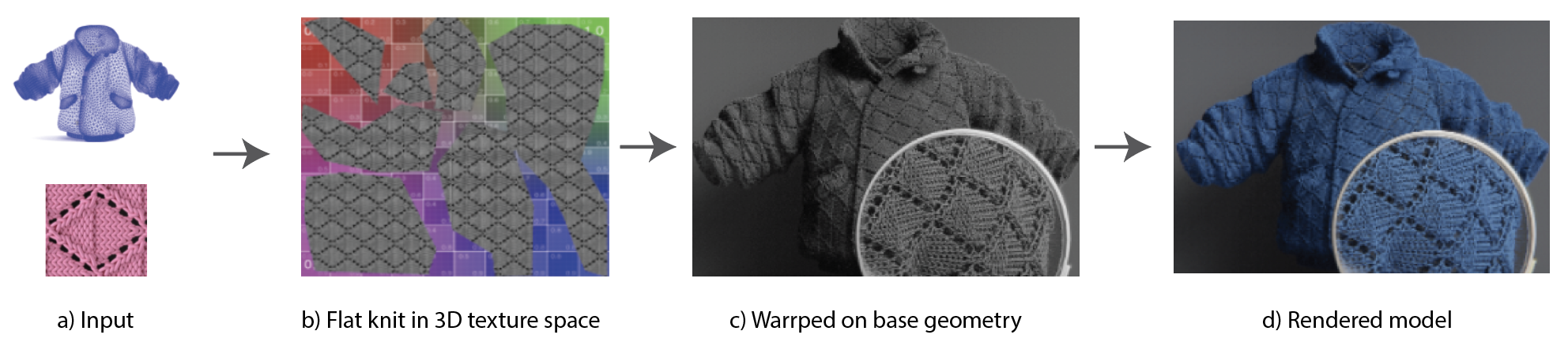}
	\caption{\textbf{Method overview.} a) The input to our system is a geometry base mesh with per-vertex UV as well as a set of curves for a tilable knitted pattern. b) The knit geometry is generated in a flat plane also referred as 3D texture space. c) Then, warped to the object by extruding out of the base geometry and forming the ply-level cloth model. d) During rendering, fiber details are added using ply-based appearance model.}
	\label{fig:pipeline}
\end{figure*}

\paragraph*{Overview.}
Our method takes as input a mesh with per-vertex UV along with a tilable relaxed yarn curves for a single knitted pattern released by Leaf et al. \cite{Leaf2018}. We first generate the knit geometry in a flat piece, throughout the paper we refer to the space where the cloth is represented as \textit{3D texture space} and then we extrude the yarns out from the base mesh to form the yarns in actual 3D object. Ply curves are generated twisting around the yarn curves given the parameters (e.g. number of plies, ply twist). Lastly, fiber geometry is added using normal/tangent and shadow mapping on the surface of ply-curve geometries same as Montazeri et al. \cite{Montazeri2020}. 

\subsection{Generating the Flat Knits}
\label{subsec:flat}
To generate the knit geometry, we first create the yarn curves in a flat piece. Given the density of the curves, number of plies and the ply twisting amount we can create the knit geometry in the 3D texture space. This method involves two stages: first tiling the knitted pattern given the number of repeating parameter as an input. Then, stitching the tilable cells together to make a connected piece of the entire knit textile.

\begin{figure}[b]
	\centering
	\includegraphics[width=0.3\textwidth]{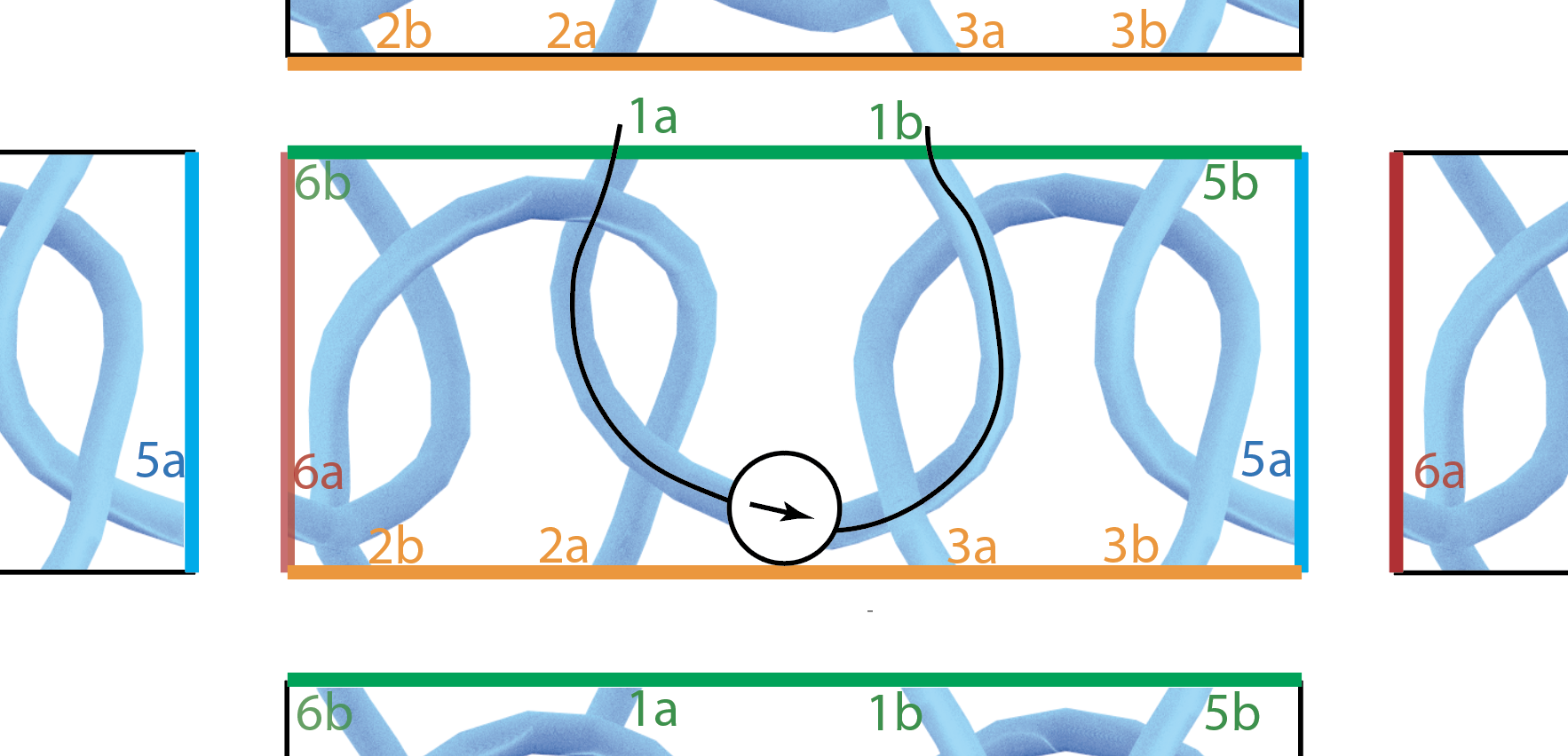}
	\caption{\textbf{Pattern tiling and stitching.} Coupling the endpoints of all the curves in the pattern cell among neighbor cells to compute how it is connected to the adjacent cells and generate the fully connected yarns. Each curve represents a node in the graph and the arrow shows the direction of the yarn which may need to be reversed based on the curves which it is connected to.}
	\label{fig:cells}
\end{figure}

\paragraph*{Pattern tiling.} Given the yarn curves of a knitted pattern, we construct a directed graph, in which each node represents a copy of the given pattern, and we refer to them as \textit{cells}. Each cell includes the curves of the pattern and each curve has two endpoints connected to the two neighbor cells. As a pre-processing step, we store a dictionary of the \textit{partner endpoints} --- the couple endpoints of a curve from one of the four neighbor cells that they are connected to. Finding the partner of an endpoint can be done by searching for the closest endpoint among all four neighbor cells (i.e. top, bottom, right, row). Therefore, all endpoints of the cell are labeled which adjacent cell and what curve of that cells it is connects to. Figure \ref{fig:cells} illustrates the labeling for each endpoint of the center cell. Since later we repeat the cell along the texture map to form the 3D geometry in a flat piece, the labels can be reused for all cells repeatedly. 

\paragraph*{Pattern stitching.} Unlike the woven fabrics, knits are usually built using a small number of yarns. Therefore, after connecting yarn curves in adjacent cells they form long yarn curves. We need the full yarns because the yarn length is required to twist the plies around the yarn center. Also, the length of the ply curves are needed for adding the fiber details. Fiber details are computed using normal/tangent as well as shadow mapping which requires the ply surface to be fully parameterized, same method as Montazeri et al. \cite{Montazeri2020}. We define a graph where each node represents each curve of every pattern cell. This establishes a degree-2 graph and we traverse the graph to generate the full curves. Figure \ref{fig:cells} also illustrates the full curve generation. Note that the order of vertices along the curves might be reversed due to the connecting edges and this is depicted using an arrow inside the curve node. So using the correct connections between the tiled cells are computed to seamlessly generate the knit topology.

\subsection{Base Geometry Transformation}
\label{ssec:transform} 
\begin{figure}[b]
    \centering
    \includegraphics[width=0.4\textwidth]{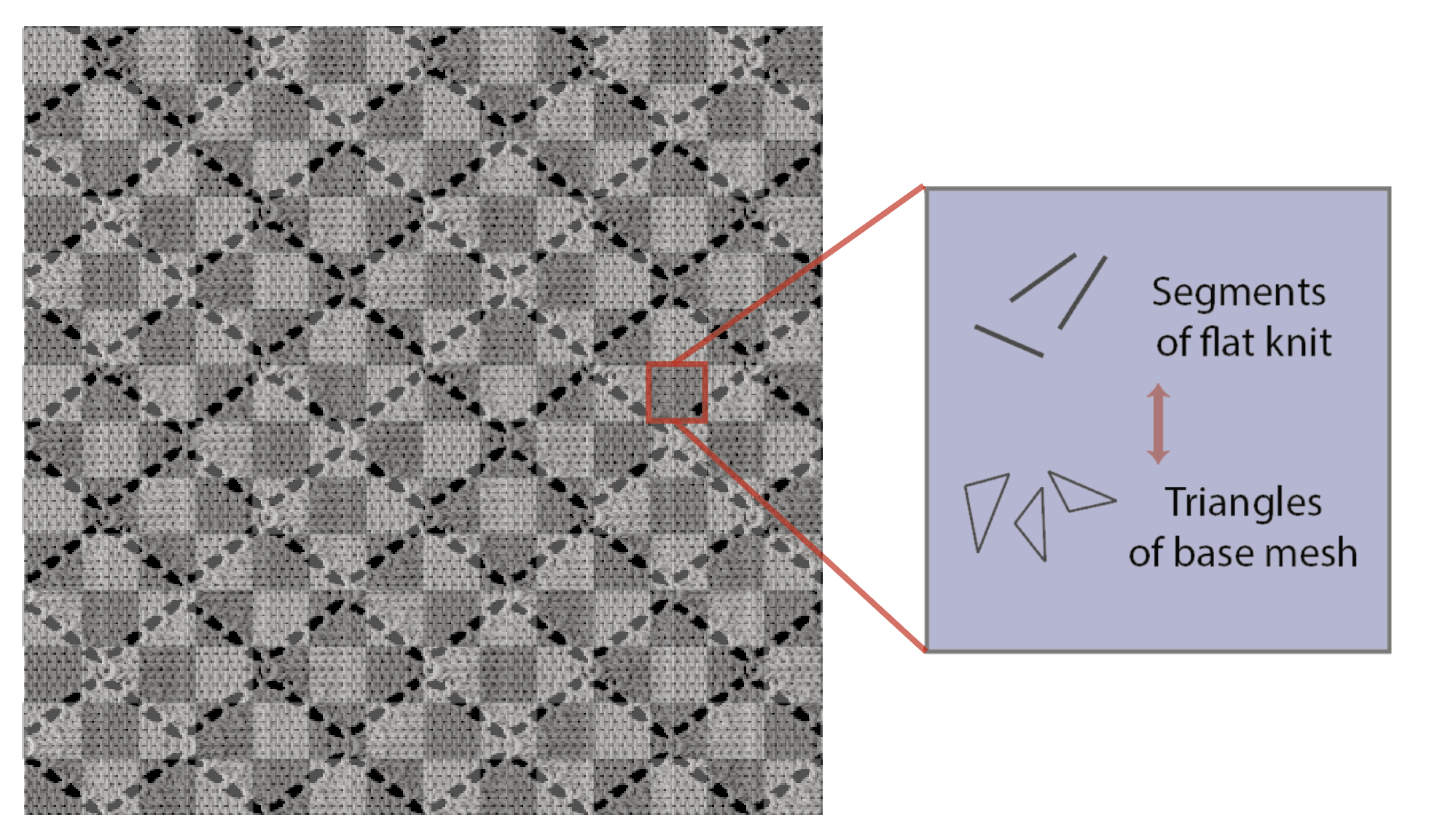}
    \caption{\textbf{Mapping grid.} The pre-computed mapping method for transferring the flat piece to the base geometry by extruding the yarns out of the base mesh. The same grid is also used to accelerate the rendering process and will be discussed in Section \ref{ssec:appearance} }.
    \label{fig:grid}
\end{figure}

Once the flat knitted fabric is generated in 3D texture space as explained in Section \ref{subsec:flat}, to transform the model into the object space, we develop an efficient mapping technique to transform the flat piece from 3D texture space to the actual geometry in object space \ref{ssec:transform}. The mapping requires a pre-computation phase with a simple implementation. The pre-computed mapping allows efficient rendering, and will be discussed in Section \ref{ssec:appearance}.
\paragraph*{Pre-computation.} We define a 2D grid to store the curve segments in 3D texture space along with their correspondent triangles from the base geometry. The 2D grid is created by subdividing the UV space into cells. Each UV cell holds two corresponding lists: first, the indices of the curve segments that lie within that particular cell, second, given the UV values of each vertex of the triangle mesh, a list of triangle indices that cover that particular cell in the base geometry. Hence, each grid cells $c$ corresponds a list of triangles in object space ($T$) to a list of curve segments in 3D texture space ($\Gamma$), with a tuple $c=(s,\tau)$ where $s(c) \in \Gamma$ and $\tau(c) \in T$. The resolution of the grid is pre-defined proportionally to the mesh density. Figure \ref{fig:grid} shows the dictionary to correspond the curve segments from the flat piece to the triangles of the base geometry. 
\paragraph*{Transformation.} To transform the flat knit as explained in Section \ref{subsec:flat} from 3D texture space to the actual object space of the base geometry, we require the following two inputs. First the UV values of all the vertices of the segment $s(c) \in \Gamma$ for the flat textile (\textit{$U$})as well as the pre-computed 2D grid which holds the correspondent triangle set $T$ and the segment set $\Gamma$ for each cell $c$.
For each $\vec{u} \in U$, given $\vec{u}$ we find the cell index $c$ as well as the set of triangles $T$ which belongs to the cell. Then using bilinear interpolation we find the exact triangle that corresponds to the segment which is $\tau(c) \in T$. Finally, Using trilinear interpolation we compute the 3D coordinate of the point on the based geometry $\vec{P} = (x, y, z)$. All vertices of the ply curve along with their normals as well as the curve length are stored which require 28 Bytes per vertex overhead in total. Algorithm \ref{algo:map} elaborates the steps further.
\begin{algorithm}[h]
\caption{Transferring the flat textile to the base geometry\label{algo:map}}

\begin{algorithmic}[1]
\raggedright
\STATE \textbf{Require:} The pre-computed 2D grid and $U$\;    
\FORALL{Every $\vec{u}  \in U $}
\STATE Find $c(\vec{u})$
\STATE Find $T$ given $c$
\STATE Find $\tau(c) \in T$ \hfill\COMMENT{using bilinear interpolation}
\STATE Compute $\vec{P}$ \hfill\COMMENT{using trilinear interpolation}
\ENDFOR 

\end{algorithmic}
\end{algorithm}
\input{appearance}

%% file: appearance.tex
\subsection{Appearance Modeling}
\label{ssec:appearance}
To faithfully re-produce the cloth appearance in ply-level, we use an aggregated BSDF model which first proposed by Montazeri et al. \cite{Montazeri2020} to model the light reflection as well as the light transmission. It is a known fact that thick yarns which happen to be more common in knit manufacturing (compared to woven fabrics), are consist of several plies. Hence, using a ply-level appearance modeling plays an important role in the knit geometry. \\
\par\textbf{Parameter fitting.} To fit the parameters of the light-scattering model automatically, we solve an inverse-rendering problem same as Montazeri et al. \cite{Montazeri2020} by minimizing the loss between the rendered results and the photo references. While they used the average intensity of the rendered image compared to the reference, we chose to leverage a learning-based method loss function to improve the accuracy of the fitting process. In addition to L2 distance between two images as loss function, a popular method is using the pre-trained neural network for imaging as VGG16 \cite{simonyan2015deep} to compute the distance between the rendering and the reference in 10\% resolution of the original image.
\subsubsection{Rendering}
\label{sssec:rendering}
We use a custom pathtracer for the rendering of our cloth models. To efficiently render the textile, we perform the raytracing in two stages, \textit{global ray-tracing } and \textit{local ray-tracing}, which are explained in the following two paragraphs, respectively. We again require the pre-computed 2D grid which holds the correspondent triangle set $T$ and the segment set $\Gamma$ for each cell $c$ same as Section \ref{ssec:transform}.
\par\textbf{Ray-surface intersection.} First, we execute a simple ray intersection with the base mesh using conventional acceleration hierarchy similar to the existing ray-tracing renderers, which returns the intersected triangle $\tau$ along with the UV value of the hit point in the object space $\vec{v}$. We refer to the \textit{ray-surface intersection} as the global stage.

\par\textbf{Ray-ply intersection.} By looking-up in the pre-computed 2D mapping grid, the cell where $\vec{v}$ belongs to is found. Given that there are already pre-stored set of segments $\Gamma$ belong to cell $c$, we compute the intersection of the ray only for those segments belonged to $c$ locally. Each segment forms a cylinder with given ply radius and the ray intersection is computed only for few segments which makes the rendering process much faster rather than finding the ray intersection with the entire segmented curves. This \textit{ray-cylinder intersection} is the local step. Once we find the hit point on the surface of a cylinder we check if it is in the trimmed section where two adjacent segments meet. In that case we compute the correct intersected point with the adjacent segment and further smooth out the normal of the hit point as explained in detail in Section \ref{ssec:detail}. The angular phase $\beta$ is needed to add the fiber details and finally, the normals and tangents are smoothed out by interpolating along the segment. This step returns $\vec{Q}$ which is the 3D position, the normal and the length of the intersected point on the surface of the ply.
\begin{algorithm}[h]
\caption{Rendering the cloth model\label{algo:render}}

\begin{algorithmic}[1]
\raggedright
\STATE \textbf{Require:} The pre-computed 2D grid and Base geometry mesh 
// Global stage: Ray-surface intersection:
\STATE Find $\tau$ and $\vec{v}$ by raytracing the base mesh

// Local stage: Ray-ply intersection:
\STATE Find $\Gamma$ belongs to $c(\vec{v})$ 
\FORALL{Every $s \in \Gamma $}
    \STATE Check intersection with the cylinder formed along the segment $s$
    \STATE Check if it hits at the joints
    \STATE Compute $\beta$ needed for fiber details
    \STATE Compute $\vec{Q}$ 
    \STATE Smooth $\vec{Q}$ along the $s$
\ENDFOR 

\end{algorithmic}
\end{algorithm}

%% file: implementation.tex
\section{Implementation Details}
\label{sec:implementation}

We implemented a custom raytracer that supports next-event estimation for both area and environment lighting. To accelerate ray-ply-intersection we perform two stages for rendering: global and local raytracing described in Section \ref{sssec:rendering}. In what follows, we provide more details on modeling process.
%
\subsection{Modeling Geometry}
\label{ssec:detail}
Given the pre-stored vertices of the piece-wise linear ply curves we compute the ray-ply intersection such that each segment of the ply curve defines a cylinder where the overall shape of the plies are round pipes. At the joints where two cylinders meet, we model the intersected ellipse perpendicular to the average tangent of the adjacent segments. To add the fiber details, 1D textures warps around the endpoint ellipses of the segment as explained in Section 3.3 of \cite{Montazeri2020}. The interpolation of the textures along the surface of the cylinder is non-trivial and was not mentioned in the previous work. In addition to the interpolation between two ellipses, to compute the angular phase of each point on the surface, the arc length of the ellipse is required. To address these challenges, we clarify our method for smoothing the fiber texture in the following. 
\par
\begin{figure}[t]
	\centering
	\hspace{-0.25in}
	\begin{tabular}{c}
    	\includegraphics[width=0.47\textwidth]{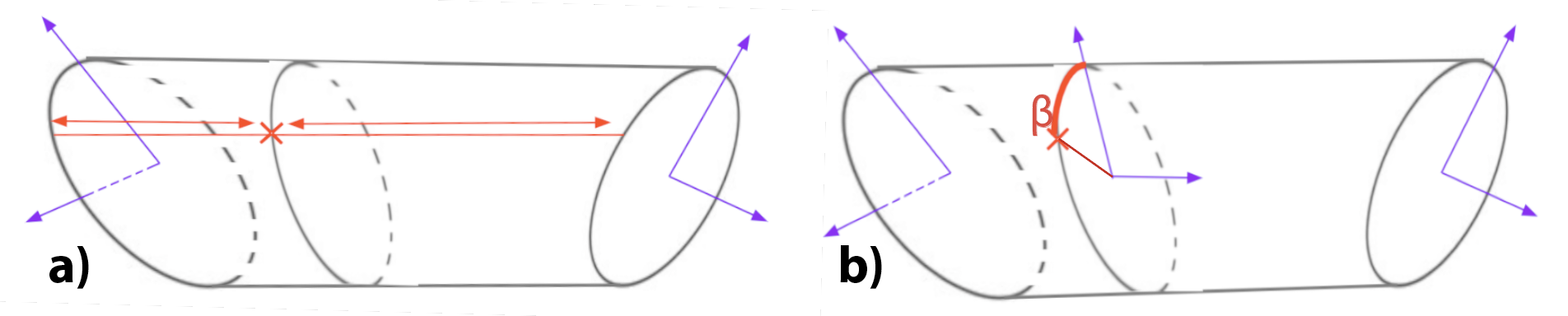}
    	%
	\end{tabular}
	\caption{(a) Illustrates the interpolation along one segment. (b) In order to map the fiber details we need to compute the angular phase around the interpolated ellipse at the intersected position, the arc length of the ellipse ($\beta$) is computed using the angle between the shading and geometry normal.}
	\label{fig:interpolation}
\end{figure}

\textbf{Smoothing fiber details} To generate continuous shading frames across the ply surface, we linearly interpolate frames of the both ends of the cylinder. However since the end caps are elliptical, computing the normals along the circumference of the ellipse is challenging, hence we compute the intersection of the line parallel to the yarn center on the surface of the cylinder with the two elliptical caps to obtain the normals of the both ends along the circumference of the ellipses. Then, tangents of the frames are updated using rotation minimizing frames \cite{Wang2008}. Figure \ref{fig:interpolation}(a) illustrates the interpolation along one segment between two elliptical endpoints. Later, in order to add the fiber details we compute the angular phase \textit{$\beta$} around the interpolated ellipse at the intersected position, the arc length of the ellipse is computed by computing the angle between the \textit{shading} and \textit{geometry} normal as shown in \ref{fig:interpolation}(b). Where shading normal is the perturbation of the geometry normal based on the 1D fiber texture. 

%% file: results.tex
\section{Results}
\label{sec:results}
In this section, we first evaluate the physical accuracy of our technique by comparing renderings to photographs of real samples (Section~\ref{ssec:validation}).
Then, we show more rendered results generated with our method for common knitting patterns on complex base geometries (Section~\ref{ssec:results}).

\subsection{Comparison with photographs}
\label{ssec:validation}
\begin{figure}[t]
    \vspace{-0.05in}
    \hspace{-0.3in}
	\centering
	\begin{tabular}{ccc}
    	& photo & Ours
    	\\
    	\raisebox{1.5\height}{\rotatebox[origin=c]{90}{Front lighting}}
    	& \includegraphics[width=0.20\textwidth]{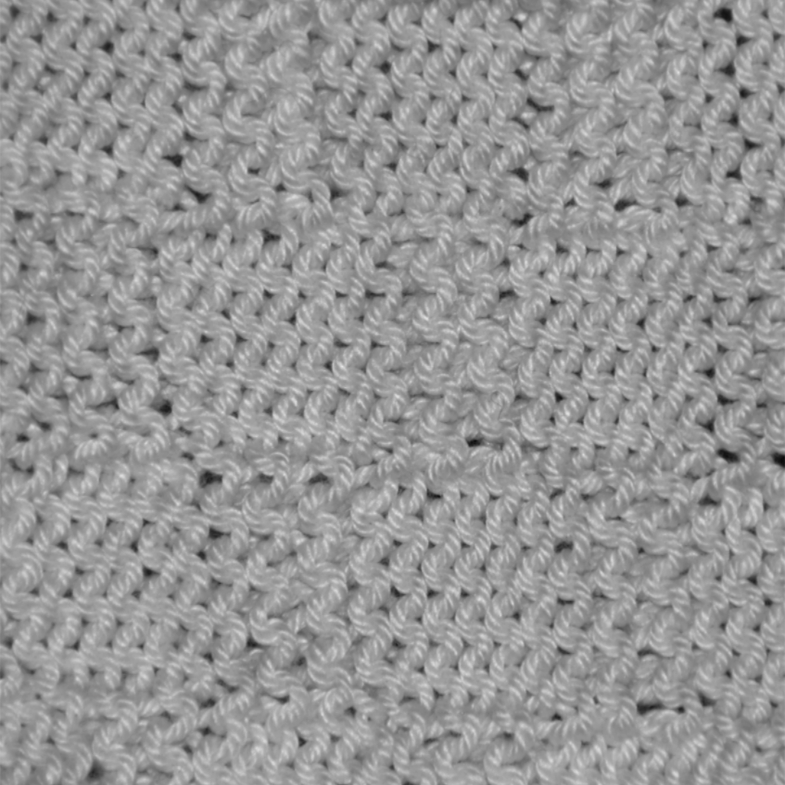}
    	& \includegraphics[width=0.20\textwidth]{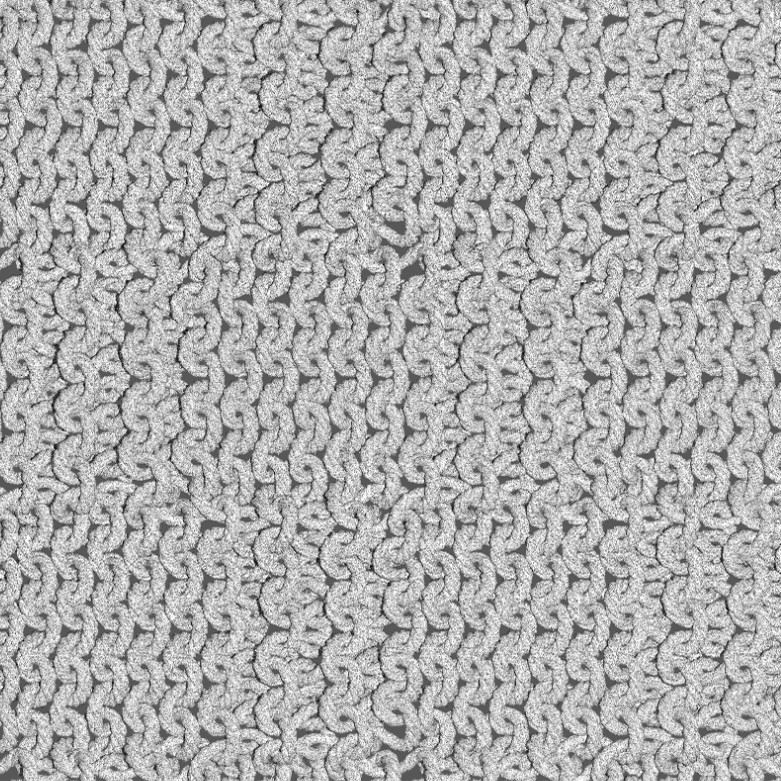}
    	\\
    	\raisebox{1.5\height}{\rotatebox[origin=c]{90}{Back lighting}}
    	& \includegraphics[width=0.20\textwidth]{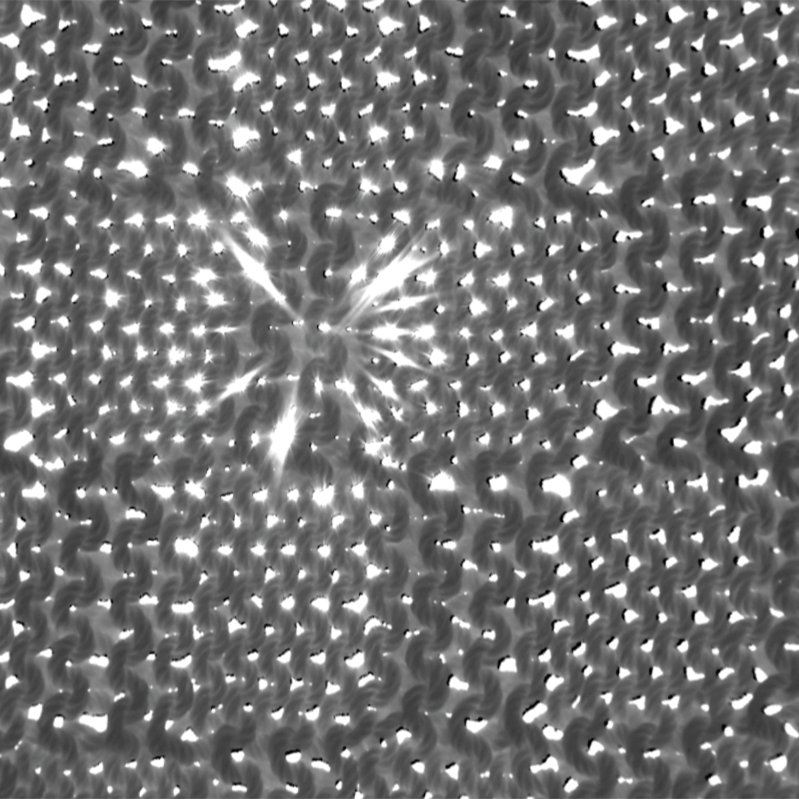}
    	& \includegraphics[width=0.20\textwidth]{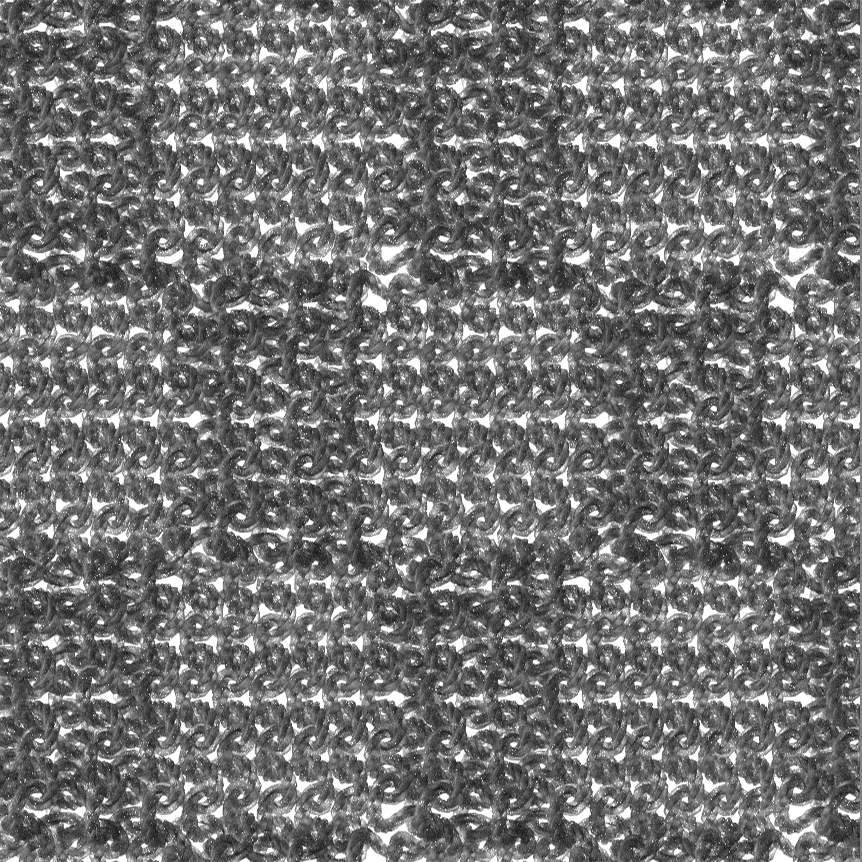}
    	\\
	\end{tabular}
	\caption{\label{fig:appearance} \textbf{Matching photographs of real sample knits with our model.} We use VGG loss to optimize the appearance parameters of the rendered image to match with the reference photos. Our system is capable of matching the appearance of the knits 
	under both front and back lighting.
	}
\end{figure}

We create virtual replica of real samples by examining each sample under a digital microscope and manually setting the geometry parameters. Then, we set the material appearance parameters automatically by solving an inverse-rendering problem. Specifically, we minimize the difference between rendered images and photographs measured using the VGG16 loss ---a learning-based metric~\cite{simonyan2015deep} with 10\% resolution of the original image--- in addition to a regular L2 distance.\\
Fig.~\ref{fig:appearance} shows photographs of a real sample (on the left) and rendered results using our system with optimized parameters (on the right).
We illuminated the cloth sample from both the front and the back using a small area light.
Both front and back lighting references are needed for the system to optimize the appearance parameters. Since having these two setups captures both reflection and transmission of light, and all of the parameters were fitted given the two references simultaneously. 

\subsection{Main Results}
\label{ssec:results}
We now show additional rendered results using our technique for complex geometries.

\begin{figure*}[b]
    \centering
    \setlength{\resLen}{0.235\textwidth}
	\addtolength{\tabcolsep}{-3pt}
	\begin{tabular}{cccc}
    	\includegraphics[width=\resLen]{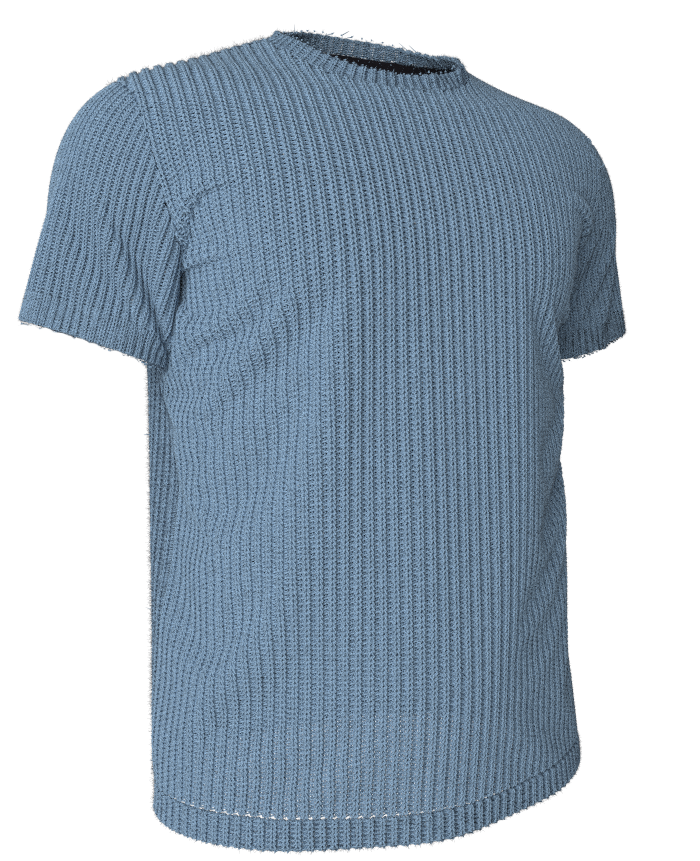}
    	& \includegraphics[width=\resLen]{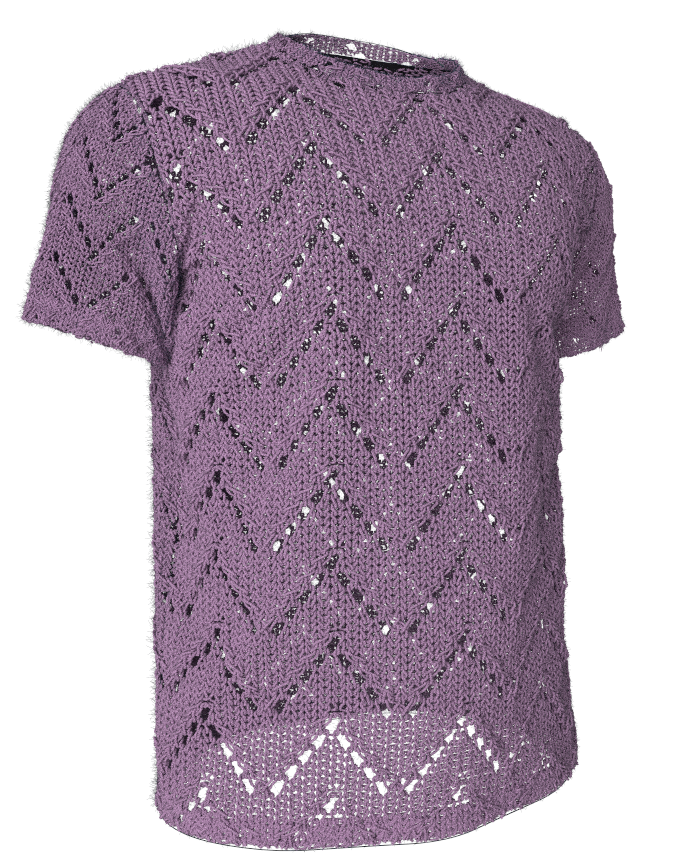}
    	& \includegraphics[width=\resLen]{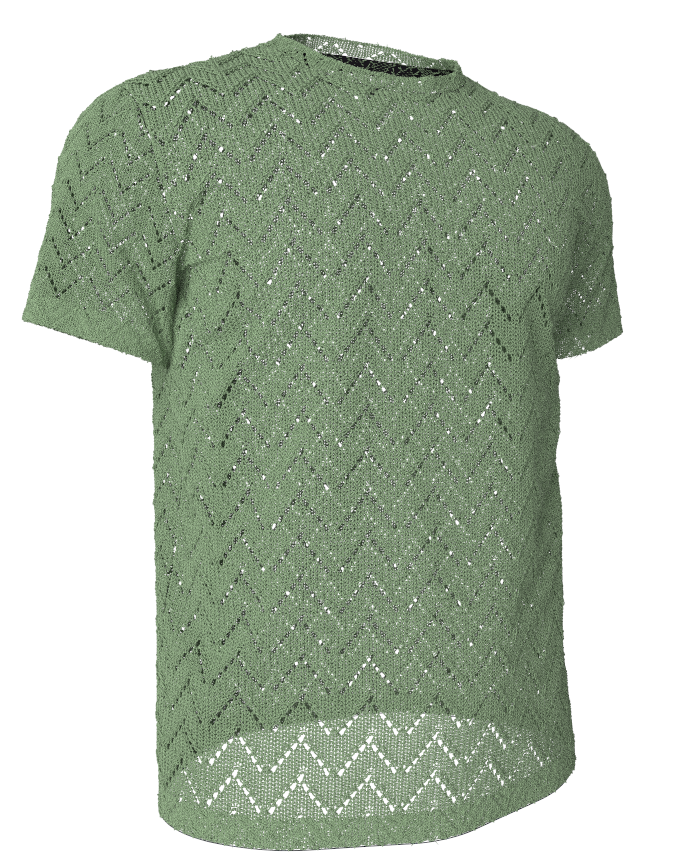}
    	& \includegraphics[width=\resLen]{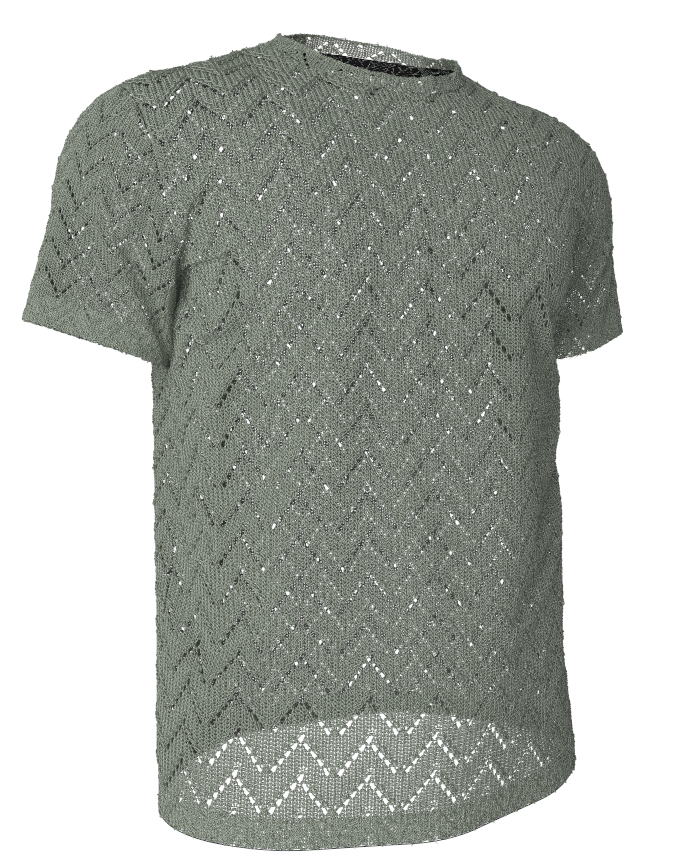}
    	\\
    	\includegraphics[width=\resLen]{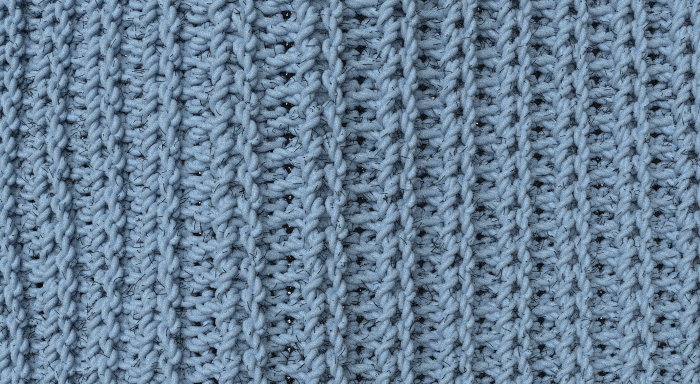}
    	& \includegraphics[width=\resLen]{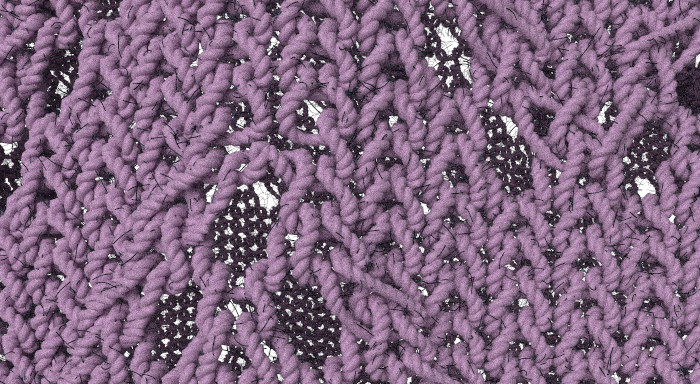}
    	& \includegraphics[width=\resLen]{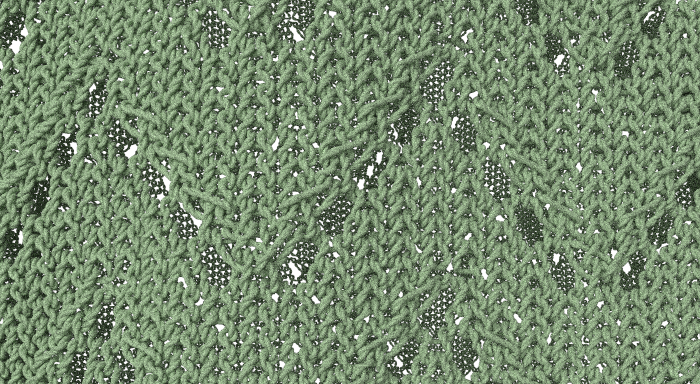}
    	& \includegraphics[width=\resLen]{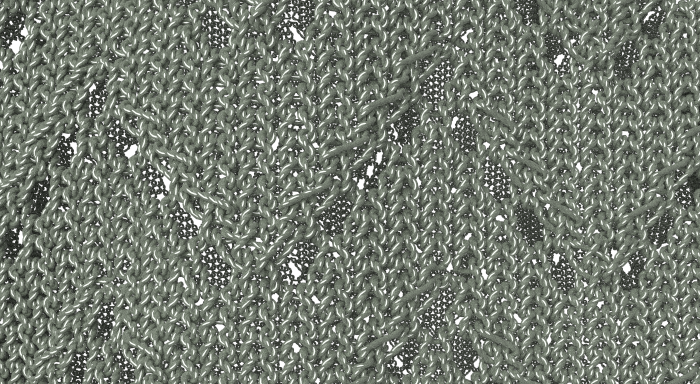}
    	\\
	\end{tabular}
	\caption{\label{fig:tshirt} 
	\textbf{Interactive editing geometry and appearance.} a) Changing the knitting pattern b) yarn density c) color d) and glossiness.
	}
\end{figure*}

Our model enjoys the flexibility to represent knitted fabrics with widely yarn structures and optical properties.
Fig.~\ref{fig:tshirt} shows such an example with several different knitting patterns, yarn densities, colors, and glossiness.
Additionally, we show in the accompanying video an example where these properties are edited interactively.

\begin{figure*}[t]
    \centering
    \setlength{\resLen}{0.235\textwidth}
	\addtolength{\tabcolsep}{-3pt}
	\begin{tabular}{ccc}
	    \includegraphics[width=0.3\textwidth]{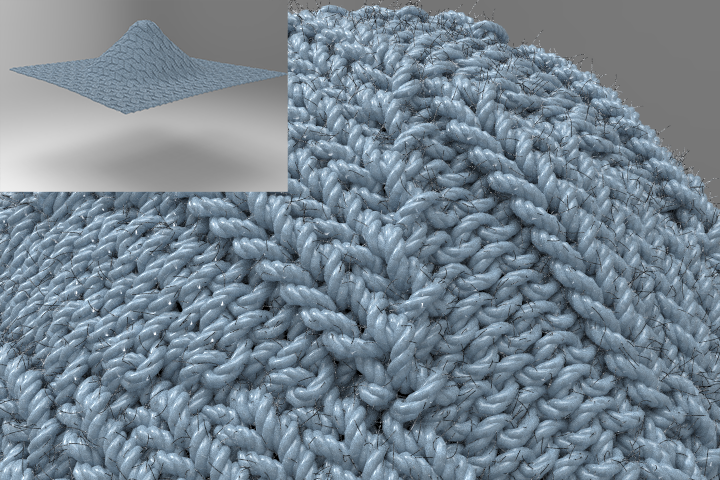}
    	& \includegraphics[width=0.3\textwidth]{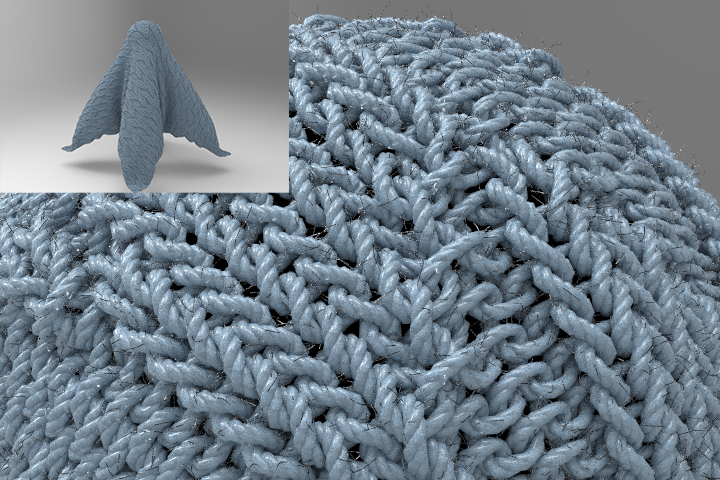}
    	& \includegraphics[width=0.3\textwidth]{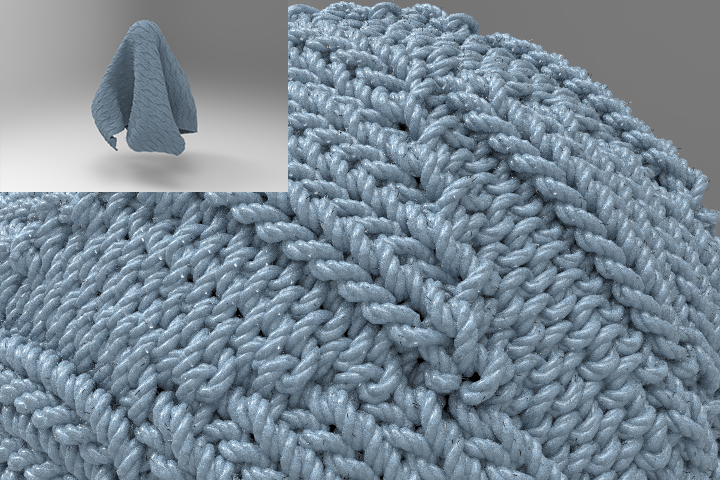}
    	\\
    	\includegraphics[width=0.3\textwidth]{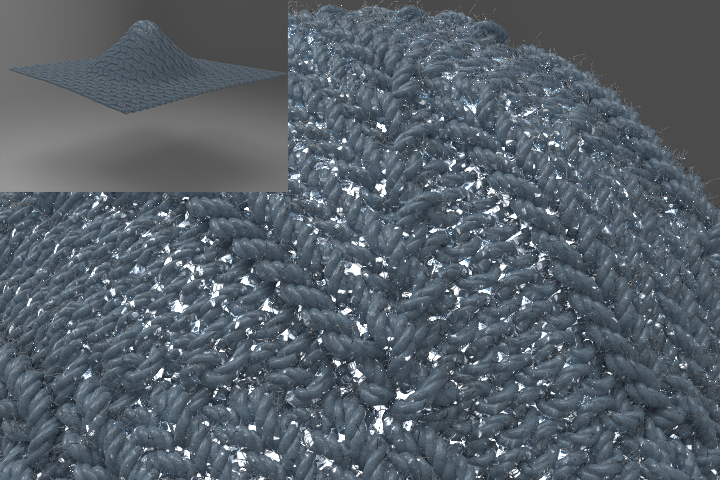}
    	& \includegraphics[width=0.3\textwidth]{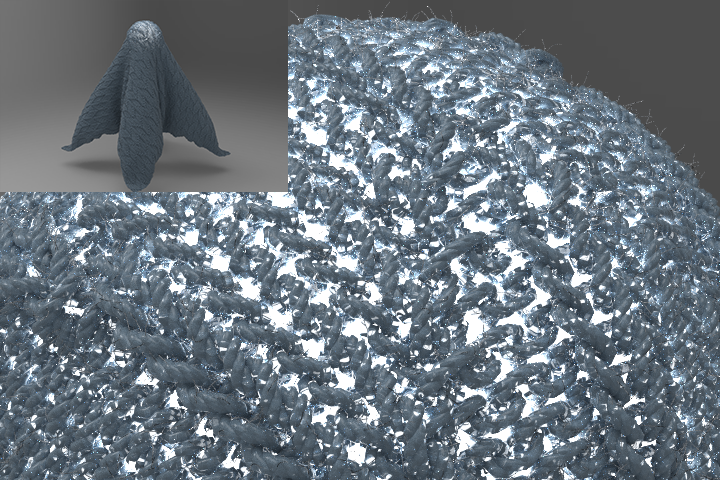}
    	& \includegraphics[width=0.3\textwidth]{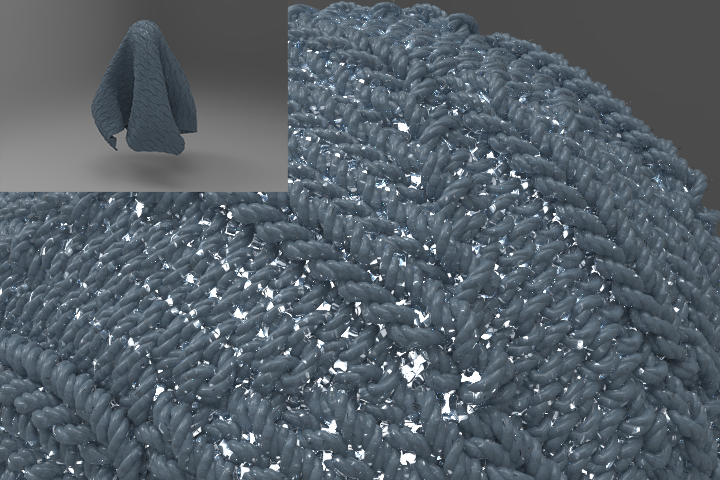}
    	\\

	\end{tabular}
	\caption{\label{fig:drape}
	    \textbf{Temporal consistency.}
	    Our model is capable to generating detailed animations of knitted fabrics with only the base mesh simulated.
	    In this example, we show four frames of an animation where a heavy piece of cloth is draped onto a spherical light (that is off in the top and on in the bottom).
	    Our technique manages to produced visually plausible deformation of yarns (e.g., the stretching of yarns when they collide with the sphere).
	}
\end{figure*}

Fig.~\ref{fig:drape} illustrates four frames extracted from an animation in which a rectangular piece of knitted fabric is draped onto a solid sphere.
The model is rendered under both front (top) and back lighting (bottom).
Our modeling technique allows detailed and physically plausible animations to be obtained with only the base mesh being simulated (that is, no yarn-level simulation).
Please see the supplementary video for the full video.

\begin{figure*}[t]
    \centering
    \setlength{\resLen}{0.225\textwidth}
	\addtolength{\tabcolsep}{-2pt}
	\begin{tabular}{cccc}
    	\includegraphics[width=\resLen]{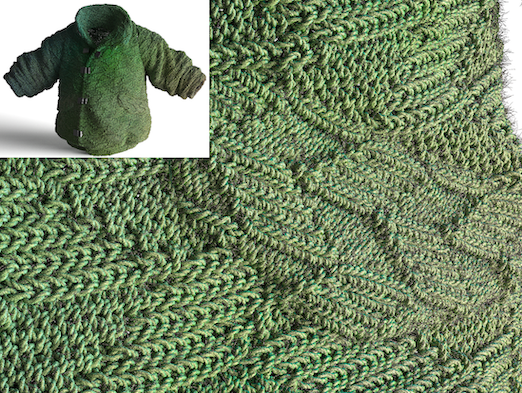}
    	& \includegraphics[width=\resLen]{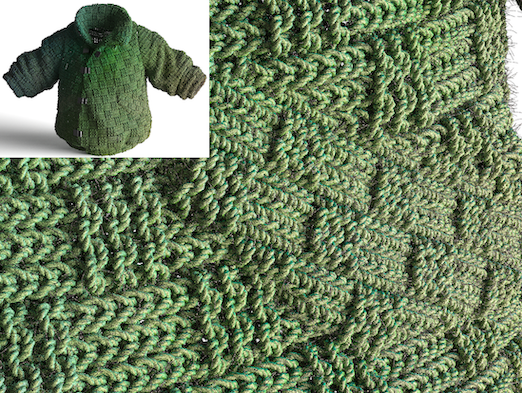}
    	& \includegraphics[width=\resLen]{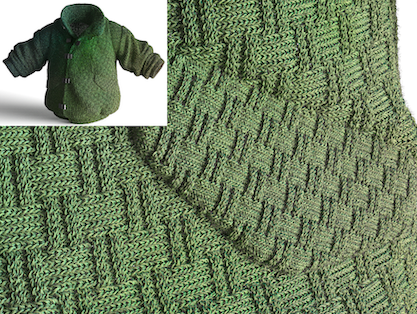}
    	& \includegraphics[width=\resLen]{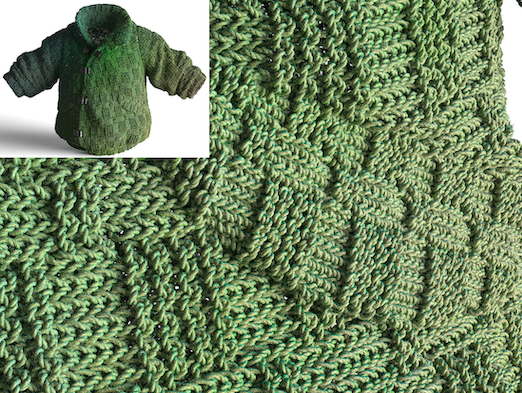}
    	\\
    	\includegraphics[width=\resLen]{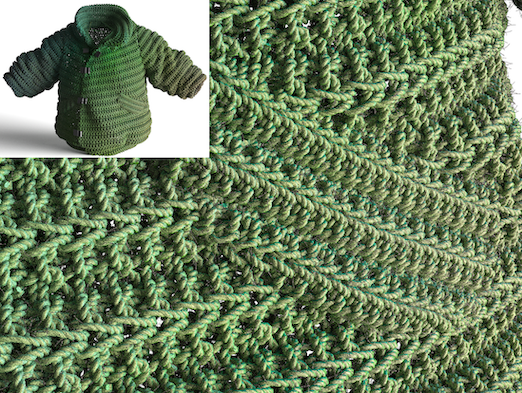}
    	& \includegraphics[width=\resLen]{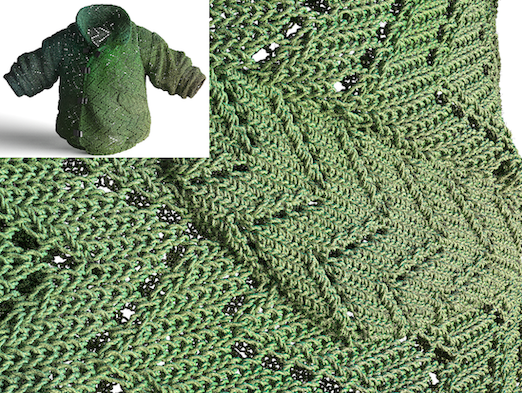}
    	& \includegraphics[width=\resLen]{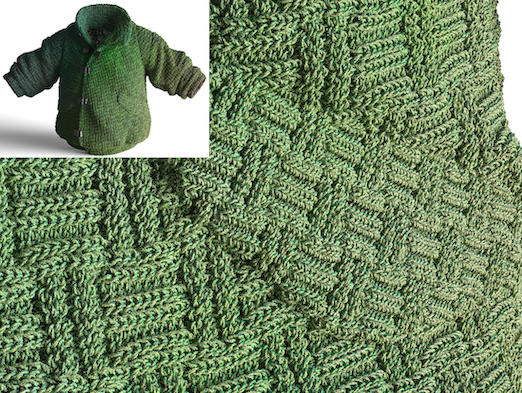}
    	& \includegraphics[width=\resLen]{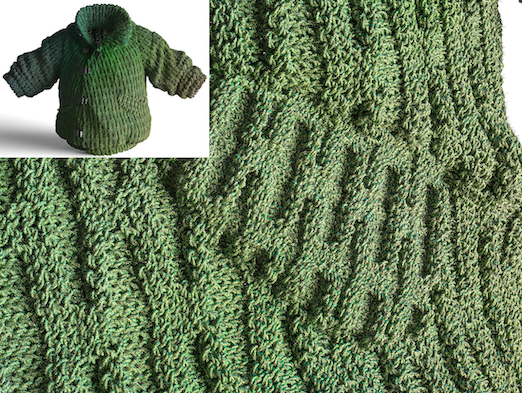}
    	\\
    	\includegraphics[width=\resLen]{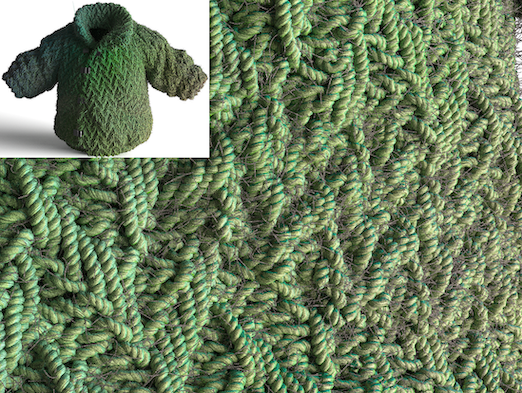}
    	& \includegraphics[width=\resLen]{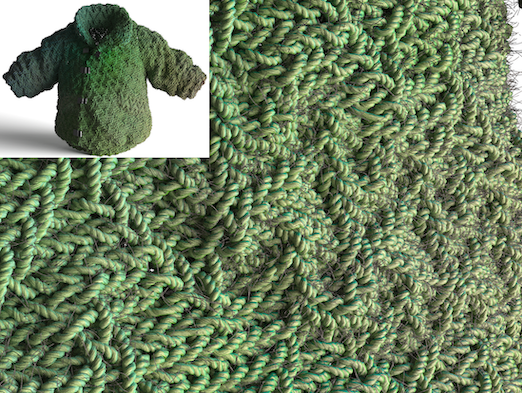}
    	& \includegraphics[width=\resLen]{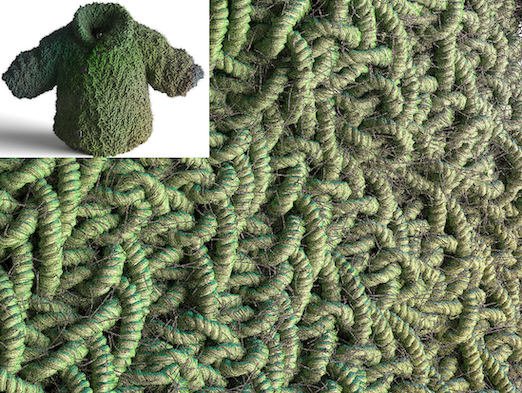}
    	& \includegraphics[width=\resLen]{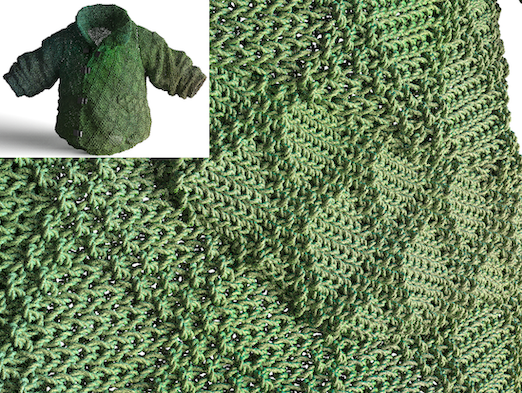}
    	\\
    	\includegraphics[width=\resLen]{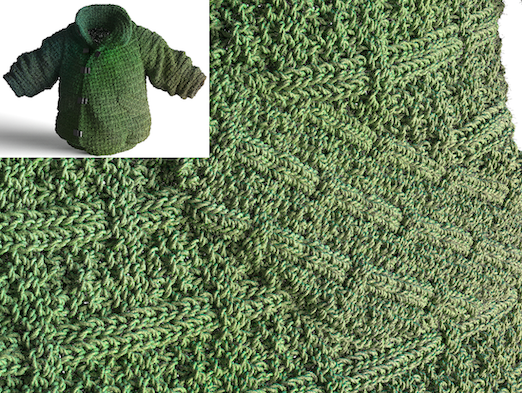}
    	& \includegraphics[width=\resLen]{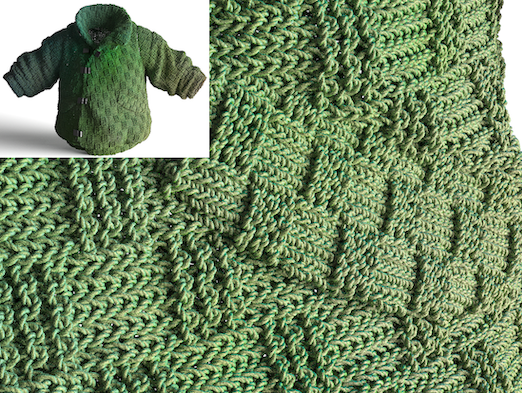}
    	& \includegraphics[width=\resLen]{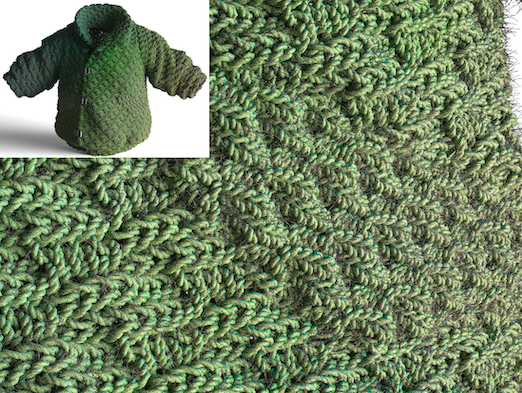}
    	& \includegraphics[width=\resLen]{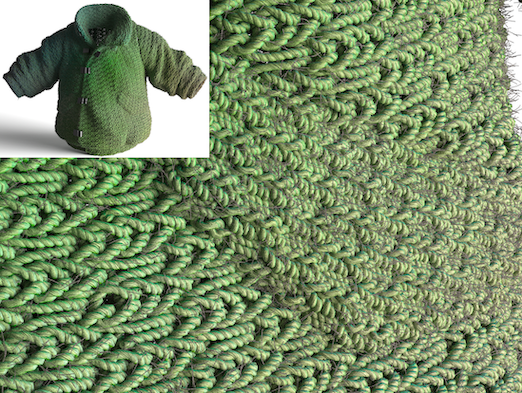}
    	\\
        \includegraphics[width=\resLen]{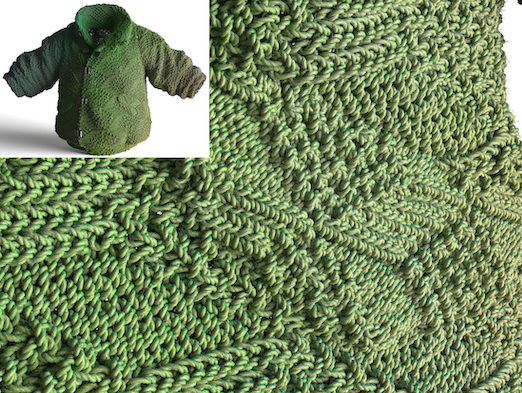}
    	& \includegraphics[width=\resLen]{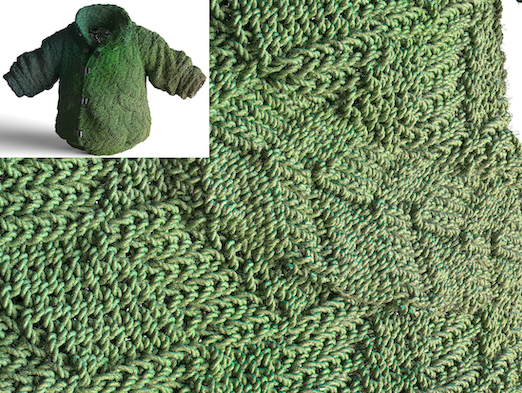}
    	& \includegraphics[width=\resLen]{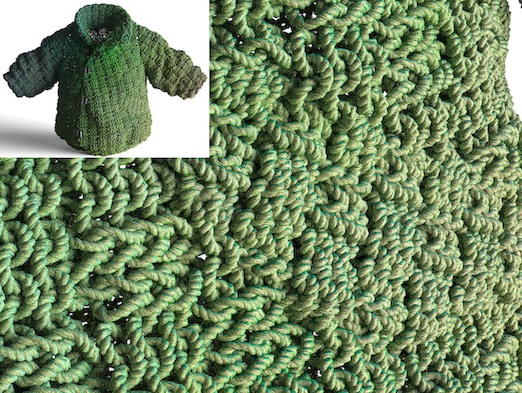}
    	& \includegraphics[width=\resLen]{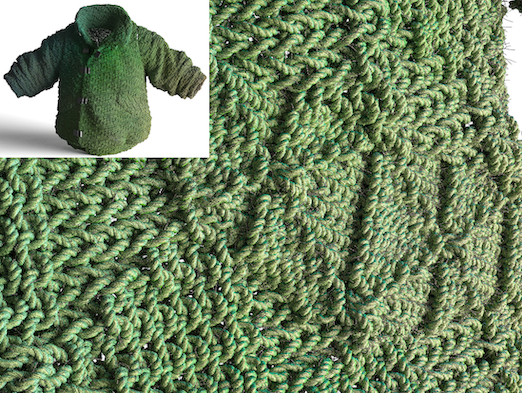}
    	\\
    	\includegraphics[width=\resLen]{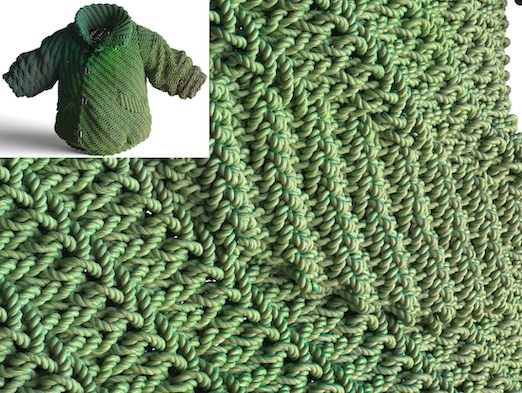}
    	& \includegraphics[width=\resLen]{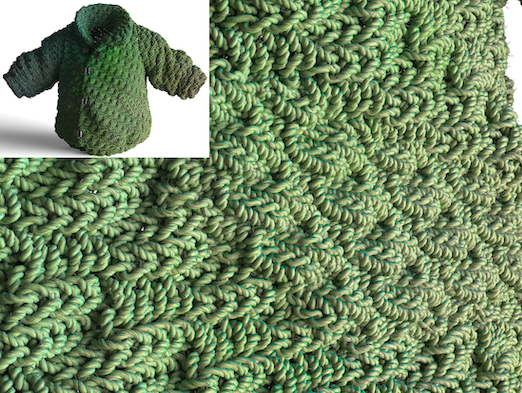}
    	& \includegraphics[width=\resLen]{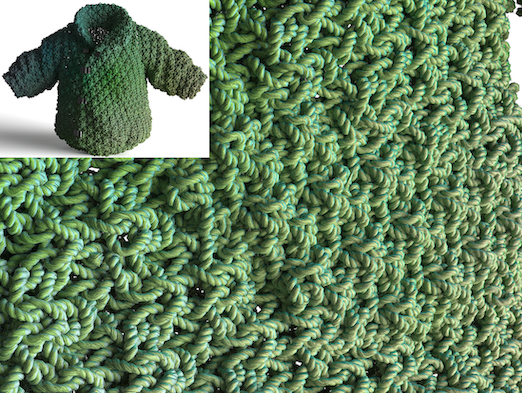}
    	& \includegraphics[width=\resLen]{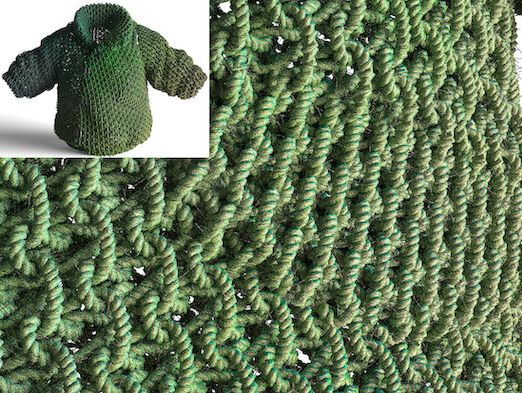}
    	\\
    	\includegraphics[width=\resLen]{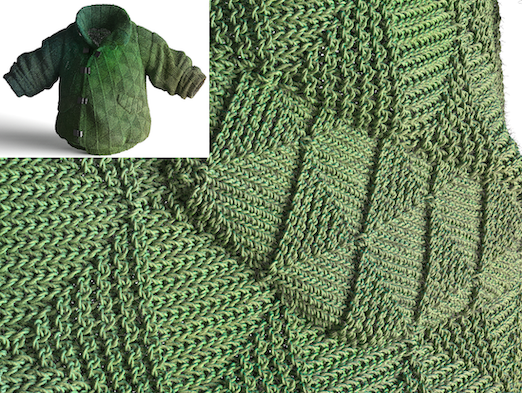}
    	& \includegraphics[width=\resLen]{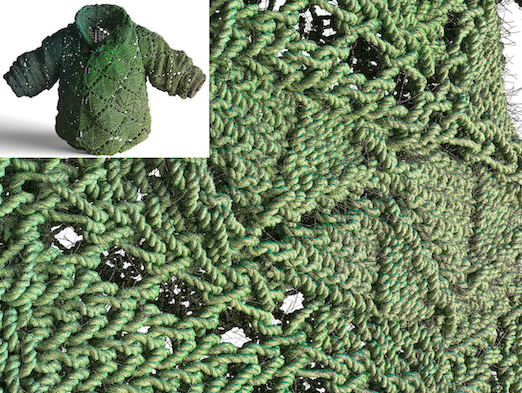}
    	& \includegraphics[width=\resLen]{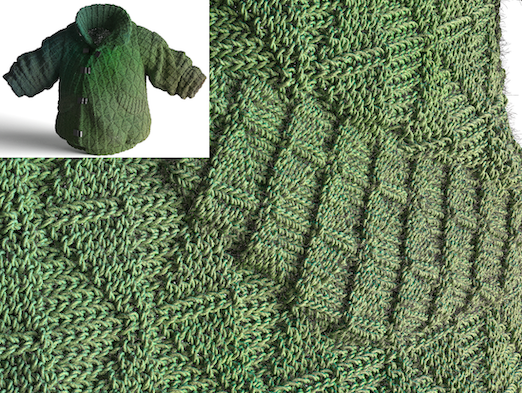}
    	& \includegraphics[width=\resLen]{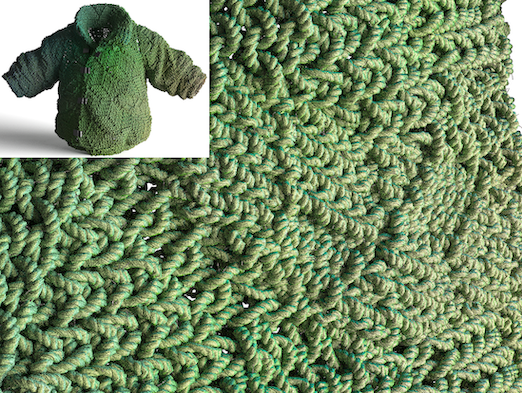}
    	\\
	\end{tabular}
	\caption{\label{fig:jacket}	Knitted fabrics with \textbf{common knitting patterns} simulated by Leaf et al. \cite{Leaf2018}.}
\end{figure*}

Lastly, in Fig.~\ref{fig:jacket} we show knitted fabrics with a wide array of patterns simulated by Leaf~et~al.~\cite{Leaf2018}. 
%
    	
%

%% file: conclusion.tex
\section{Discussion and Conclusion}
\label{sec:conclusion}
\paragraph*{Limitations and future work.}
Our technique only works for limited tilable patterns that are pre-relaxed (using yarn-level simulation).
Extending our system to support arbitrary knitting patterns is an interesting topic for future research.

Additionally, developing multi-resolution or level-of-detail versions of our geometric and appearance models could enable interactive high-fidelity rendering of knitted fabrics for VR/AR applications.

\paragraph*{Conclusion.}
We have presented a new technique that efficiently models knitted fabrics with tilable patterns and arbitrary base geometries.
Our model depicts a knitted fabric at the ply level---similar to the work by Montazeri et al. \cite{Montazeri2020}---and add fiber-level details using tangent and normal mapping.
Taking as input a base mesh and yarn curves of a tilable knitting pattern, we introduced a new pipeline that build our ply-based model automatically.
Additionally, we developed a system to efficiently render our models.

We demonstrate qualitatively that our models to reproduce the appearance of real-world knitted fabrics under both front- and back-lit configurations.
Further, our model allows knitted fabrics with fiber-level details to be animated in a physically plausible fashion by only simulating its base geometry.